\documentclass{extarticle}
\usepackage{geometry}
\usepackage[T1]{fontenc}

\usepackage{amsmath,amssymb,amsthm}

\newtheorem{proposition}{Proposition}
\newtheorem{lemma}{Lemma}
\newtheorem{theorem}{Theorem}
\theoremstyle{definition}
\newtheorem{definition}{Definition}

\usepackage[hidelinks]{hyperref}
\hypersetup{
		pdfencoding=auto, 
		psdextra,
		colorlinks=true,
		citecolor=green!40!black,
		linkcolor=red!50!black,
		urlcolor=blue!80!black,
	}
\usepackage[utf8]{inputenc}
\usepackage[english]{babel}
\usepackage{xspace}
\usepackage{microtype}
\usepackage{tikz}
\usetikzlibrary{calc}

\usepackage{enumitem}
\usepackage{natbib}
\renewcommand{\cite}[2][]{\citeyearpar[#1]{#2}}

\usepackage{framed}
\usepackage{tabularx}

\newlength{\RoundedBoxWidth}
\newsavebox{\GrayRoundedBox}
\newenvironment{GrayBox}[1]%
   {\setlength{\RoundedBoxWidth}{.93\columnwidth}
    \def\boxheading{#1}
    \begin{lrbox}{\GrayRoundedBox}
       \begin{minipage}{\RoundedBoxWidth}}%
   {   \end{minipage}
    \end{lrbox}
    \begin{center}
    \begin{tikzpicture}%
       \node(Text)[draw=black!0,fill=white,rounded corners,inner sep=2ex,text width=\RoundedBoxWidth]
             {\usebox{\GrayRoundedBox}};
        \coordinate(x) at (current bounding box.north west);
        \node [draw=white,rectangle,inner sep=3pt,anchor=north west,fill=white]
        at ($(x)+(6pt,.75em)$) {\boxheading};
    \end{tikzpicture}
    \end{center}}

\newenvironment{defproblemx}[1]{\noindent\ignorespaces%
                                \FrameSep=6pt%
                                \parindent=0pt%
                \vspace*{-1em}
                \begin{GrayBox}{#1}%
                \begin{tabular*}{\columnwidth}{@{\hspace{.1em}} >{\itshape} p{1.5cm} p{0.74\columnwidth} @{}}%
            }{
                \end{tabular*}%
                \end{GrayBox}%
                \ignorespacesafterend
            }

\newcommand{\defProblemQuestion}[3]{%
  \begin{defproblemx}{#1}
    Input: & #2 \\
    Question: & #3
  \end{defproblemx}
}

\def\ve#1{\mathchoice{\mbox{\boldmath$\displaystyle\bf#1$}}
    {\mbox{\boldmath$\textstyle\bf#1$}}
    {\mbox{\boldmath$\scriptstyle\bf#1$}}
    {\mbox{\boldmath$\scriptscriptstyle\bf#1$}}}

\newcommand\veb{{\ve b}}

\newcommand\vem{{\ve m}}

\newcommand\vex{{\ve x}}
\newcommand\vey{{\ve y}}
\newcommand\vez{{\ve z}}
\newcommand\vezero{{\ve 0}}

\DeclareMathOperator{\poly}{\operatorname{poly}}

\newcommand{\hy}{\hbox{-}\nobreak\hskip0pt}

\newcommand{\problemName}[1]{\textsc{#1}}
\newcommand{\EEFAllocationProblem}{\problemName{EEF--Allocation}\xspace}
\newcommand{\EFProblem}{\problemName{\ensuremath{\mathcal{E}}\hy{}Efficient \ensuremath{\mathcal{F}}\hy{}Allocation}\xspace}
\newcommand{\PILP}{\problemName{PILP}}

\def\instance{\ensuremath{I}}

\title{\textbf{High-Multiplicity Fair Allocation Using Parametric Integer Linear
Programming}}
\author {
  Robert Bredereck,\textsuperscript{\rm 1,4}
  Andrzej Kaczmarczyk,\textsuperscript{\rm 2,4}\\
  Du{\v{s}}an Knop,\textsuperscript{\rm 3,4}
  Rolf Niedermeier\textsuperscript{\rm 4}
  \vspace{0.3cm}\\
	\begin{minipage}{\textwidth}
		\centering
    \textsuperscript{\rm 1} Institut für Informatik, TU Clausthal,
	  Clausthal-Zellerfeld, Germany\\
    \textsuperscript{\rm 2} AGH University, Krak\'{o}w, Poland\\
    \textsuperscript{\rm 3} Czech Technical University in Prague, Prague, Czech
	  Republic\\
    \textsuperscript{\rm 4} Technische Universit\"at Berlin, Berlin, Germany
  \end{minipage}\vspace{0.3cm}\\
	\begin{minipage}{\textwidth}
		\centering\small
		\texttt{robert.bredereck@tu-clausthal.de, andrzej.kaczmarczyk@agh.edu.pl}\\
		\texttt{dusan.knop@fit.cvut.cz, rolf.niedermeier@tu-berlin.de}
  \end{minipage}
}
\date{}

\begin{document}
\maketitle

\begin{abstract}
 Using insights from parametric integer linear programming, we
 improve the work of Bredereck et al.~[Proc.~ACM~EC~2019] on high-multiplicity
 fair allocation. Answering an open question from their work,
 we proved that the problem of finding envy-free Pareto-efficient
 allocations of indivisible items is fixed-parameter tractable
 with respect to the combined parameter ``number of agents'' plus ``number
 of item types.''
 Our central improvement, compared to their result, is to break the condition
 that the corresponding utility and multiplicity values have
 to be encoded in unary, which is required there. Concretely, we show that, while preserving
 fixed-parameter tractability, these values can be
 encoded in binary. Thus, we substantially expand the range of feasible utility
 and multiplicity values.\bigskip

 \noindent \textbf{Keywords:} fixed-parameter tractability; fixed dimension; fair
 division; efficient envy-free allocation; indivisible items
\end{abstract}

\section{Introduction}
Fairly allocating (indivisible) items~\citep{BCM16} is a key issue in a world of
limited resources, which is, for instance, reflected by multiple application
contexts such as distributing food by food banks~\citep{Wal15}, university
course assignment problems~\citep{Bud11}, or sharing computing
resources~\citep{GZHKSS11}. In recent decades, studying fair allocation issues
through the computational lens or, more generally, applying computer science
toolbox~\citep{Walsh21} proved useful in advancing our knowledge of how to deal
with finding desirable allocations. Examples include popular tools such as the
Adjusted Winner Procedure~\citep{BT96} or the web platform
spliddit.org~\citep{GP14} to name a few.

In this work, we focus on the so-called ``high-multiplicity fair allocation''
scenario in which various item types come in multiple copies.
To understand important facets of our research contribution, let us, however,
become more precise on the studied problem and the most relevant existing
results.

We consider a set of item types, each coming with the number of
actual items of this type, and a set of agents who report their non-negative utilities over
each item type. An allocation of items is an assignment of disjoint sets of the items,
called bundles, to the agents. In our work we first focus on one of the
most prominent fairness concepts which
is \emph{envy-freeness}. It considers an allocation as \emph{fair}
if there is no agent that would prefer a bundle of any other agent over
her own one. However, it is trivial to achieve envy-freeness by giving every
agent an empty bundle. To circumvent this issue, several
``efficiency'' measures of
allocations have been proposed. A very important one, \emph{Pareto-efficiency},
requires that for an efficient allocation there exists no other allocation that
is preferred by at least one agent and, at the same time, does not make any
agent worse off.
Combining the aforementioned concepts together, we end up with so-called
\emph{envy-free Pareto-efficient allocations} on which we mostly focus in this paper.

Finding envy-free Pareto-efficient allocations is a computationally
very hard problem. For instance, the corresponding decision problem
is $\Sigma_2^P$-complete for general utilities~\citep{BL08}. The hardness holds
even for (positive) additive utilities~\citep{KBKZ09}---here, the utility that
an agent gets from a bundle is a sum of utilities that this agent reports for
every item in the bundle. This model, due to its simplicity is
frequently assumed in the scientific social choice
literature~\citep{BCULP16,BT96,Rothe15} and also forms an important part of
experimental studies~\citep{BFKKR21,DGKPS14}. Notably, practically relevant
tools (like the Adjusted Winner Procedure and the web platform\footnote{The
spliddit.org webpage is currently (April 2023) unavailable. However, a
github repository with the software is available at
\url{https://github.com/jogo279/spliddit}.} spliddit.org~\citep{GP14}) make use
of additive utilities too.

Motivated by a high practical relevance of the problem of finding envy-free
Pareto-efficient allocations assuming additive utilities, Bliem et
al.~\cite{BBN16} studied its fine-grained computational complexity providing
several parameterized-tractability results. However, they left
open
a question whether the subject problem is fixed-parameter tractable with
respect to the (combined) parameter~``number of agents plus number of item types.''\footnote{Technically, the open
question was formulated for the parameter~$n + u_{\text{diff}}$,
where $u_{\text{diff}}$~denotes the number of different values in the utility
functions. This parameter can easily be seen to be equivalent to our parameter~$n+m$
in terms of fixed-parameter tractability.
Note that Bliem et al.~\cite{BBN16}
used the variable name~$m$ for the number of items and showed fixed-parameter
tractability for this parameter.}
The question was then answered partially positively (with the restriction of unary encoded item multiplicities and utilities)
in the work of Bredereck~et~al.~\cite{BredereckKKN19}.

\paragraph{Our Contribution}
Our main contribution is to strengthen the previous result of
Bredereck~et~al.~\cite{BredereckKKN19} by providing an algorithm offering better
computational complexity lower-bound guarantees for finding envy-free
Pareto-efficient allocations. To this end, applying techniques from parametric
integer linear programming, we generalize their fixed-parameter tractability result
regarding the parameterization by the number of agents and the number of
item types. Specifically, we relax the requirement of unary encoded item
multiplicities and utilities thereby allowing binary encodings.

Our result expands the range of values that we can deal with efficiently in the
case of small numbers of agents and item types. Arguably, the case is quite
relevant in practice, as all scenarios in the experimentally
studied~\citep{BFKKR21} data from spliddit.com~\citep{GP14} mostly considered at
most~$8$~agents and $10$~item types (with very few instances having at
most~$15$~agents and~$30$~item types). Additional examples could include stock
inheritance. Here, a portfolio consisting of around $30$ companies (item types)
is commonly advised by the experts. As the portfolio value grows, the number of
share units (item multiplicities) of each company to share in the inheritance
process can easily reach thousands. For such scenarios, algorithms guaranteeing
fixed-parameter tractability for binary encoding of item multiplicities are a
better bet for obtaining practically relevant running times than algorithms
assuming unary encoding.

Furthermore, similarly to their result, our technique is applicable to a broad
family of allocation problems emerging from different desiderata chosen to
represent fairness (e.g.,\ (group) envy-freeness, (group) envy-freeness up to
one good, (group) envy-freeness up to any good, maximin share) and efficiency
(e.g.,\ completeness, welfare maximization, group Pareto-efficiency).

Overall, providing our result, we mainly contribute to the improvement of 
algorithmic tools allowing for searching \emph{provably} fair and efficient
allocations of indivisible items. Notably, our technique does not only allow
for answering the question of the existence of fair and efficient allocations
but it outputs such an allocation if it exists.

\subsection{Related Work}

Our work brings together the two worlds of fair allocations and parameteric
Integer Linear Programs. Hence, we split the discussion of the related work
into two parts organized thematically. We note that due to a flurry of
literature dealing with fair allocations, we only focus on the works most
relevant to ours.

\paragraph{Efficient and Envy-free Allocations of Indivisible Resources.}
Bouveret and Lang~\cite{BL08} were the first to study the computational complexity of computing
Pareto-efficient and envy-free allocations of indivisible items in a systematic way.
Their findings include $\Sigma_2^P$-completeness for the so-called monotonic dichotomous preferences
as well as NP-hardness and polynomial-time solvability for several special cases.
Most relevant to our setting with additive utility-based preferences, they showed that
even if there are just two agents or if every agent assigns either utility value~$0$ or~$1$
to each item, the problem of finding a Pareto-efficient and envy-free allocation remains NP-hard.
Moreover, de Keijzer et al.~\cite{KBKZ09} showed that $\Sigma_2^P$-completeness even holds
for positive additive preferences.
Bliem et al.~\cite{BBN16} analyzed the parameterized complexity, showing that the problem becomes
tractable for the parameter ``number of items'' and various special settings but remains intractable
for the parameter ``number of agents.'' 

Multiple approaches have been developed to relax fairness concepts in order to circumvent computational
intractability as well as possible non-existence of Pareto-efficient and envy-free allocations.
For instance, Lipton et al.~\cite{LMMS04} considered the concept of envy-freeness up to one good (EF1).
Herein, every agent compares its bundle with the bundles of all other agents
and she is envious if any other bundle minus the most valuable item in there is better
than her own bundle.
Further studied concepts include envy-freeness up to any good (EFX)~\citep{CKMPSW16,PM18}, minimum envy~\citep{LMMS04},
group envy-freeness, group Pareto-efficiency~\citep{AW18}, or graph
envy-freeness~\citep{AKP17,BQZ17,BKN22,ABCGL18}.
Amanatidis et al.~\cite{ABM18} provide a comparison of approximate or relaxed fairness notions.

Caragiannis et al.~\cite{CKMPSW16} showed how to compute an allocation that
maximizes Nash welfare and thus yields Pareto-efficiency and EF1.
Barman et al.~\cite{BKV18} improved this result and developed an algorithm that
computes an allocation that is Pareto-efficient and EF1 with pseudo-polynomial
running time (being polynomial in the number of agents, the number of items,
and the maximum utility).
While a round-robin allocation of items can be used to obtain a complete EF1
allocation in polynomial time when all items have positive utilities, Aziz et
al.~\cite{ACI18,ACIW19} have argued that this procedure fails when items may
have negative utilities.
Leaving the complexity of computing Pareto-efficient and EF1 allocation (when
negative utilities are allowed) open, they showed that a complete EF1
allocation can be found in polynomial time even when items with negative
utilities are present.

The setting of high-multiplicity items (where items come in multiple copies)
deserves a separate treatment. Copies of items played an important role in the
seminal work of Budish~\cite{Bud11}. However, there each agent's bundle was
assumed to have to at most a single copy of a given resource (this follows from
the fact that the author was focusing on an assignment problem, like assigning
students to courses). Later, Gafni~et~al.~\cite{GHLT21} proposed a framework
for studying the existence of EFX allocations in this model. The setting where an agent can
obtain more resources of the same type was, to the best of our knowledge, first
considered by~Bredereck~et~al.~\cite{BredereckKKN19} (on whose work we improve
on). They establish a theoretical ILP-based framework for computing various
types of efficient and fair allocations. The framework was later implemented
and tested on real-data by~Bredereck~et~al.~\cite{BFKKR21}. Implicitly, the
high-multiplicity setting is also present in the work of~Eiben
et~al.~\cite{EGHO23}. They study parameterized complexity of finding graph
envy-free allocations considering a parameterization (among others) by the
number of item-types. The high-multiplicity regime has also been reinvented
by~Gorantla~et~al.~\cite{GMV23} in the context of studying the conditions under
which EF1 allocations exist.

\paragraph{Parametric ILP Aplications.}
Eisenbrand and Shmonin~\cite[Theorem~4.2]{EisenbrandS08} gave an algorithm
that, if the number of variables is fixed, solves the given instance of
Parametric ILP (\PILP{}) in polynomial time (we formally define \PILP{} in the
Preliminaries). K{\"o}ppe et al.~\cite{KQRT10} showed that one can express the
negation of bilevel integer programs (a family of certain linear programs) as
\PILP{} and used the result of Eisenbrand and Shmonin to obtain polynomial-time
solvability of bilevel integer programs in some restricted cases.

To the best of our knowledge, Crampton et al.~\cite[Corollary
2.2]{CramptonGKW19} were the first to give an ``interpretation'' of the result
of Eisenbrand and Shmonin~\cite{EisenbrandS08} in terms of parameterized
complexity analysis. More specifically, they showed membership in the complexity
class~\textsf{FPT}, that is, they showed a running time~$f(p,n) \cdot |\instance|$ for an
instance~$\instance$ of \PILP{}
provided that the coefficients of the matrix~$A$ are encoded in unary. Using
this result
Crampton et al.~\cite{CramptonGKW19}
initiated the parameterized study of the so-called \emph{resiliency} problems~
(such as the \textsc{Resiliency Closest String} problem).

Knop et al.~\cite{KnopKM18} used the interpretation of Crampton et
al.~\cite{CramptonGKW19} to solve a decade-long-standing open question of
\textsf{FPT}-membership of a variant of the \textsc{Bribery} problem in the
field
of elections manipulation.
Recently, Bredereck et al.~\cite{BredereckKKN19} also used the
interpretation of Crampton et al.~\cite{CramptonGKW19} in the context of fair
allocation.
More specifically, they
showed~\citep[Corollary~5]{BredereckKKN19} that finding a fair and efficient
allocation is fixed-parameter tractable for few agents and few item types. The
result holds for numerous different concepts of fairness and efficiency.
Yet, their result holds only when the maximum utility value an agent assigns to
an item type and item multiplicities are encoded in unary.
As we shall shortly see, we are improving upon this result by allowing item
multiplicities to be encoded in \emph{binary}.

\subsection{Organization}
In the following Section~\ref{sec:preliminaries}, we first give necessary notation and formal
preliminaries regarding allocations, parameterized complexity, and
parameterized integer linear programs. Then, in Section~\ref{sec:preparation},
we lay foundations for proving our main result by presenting a convenient
interpretation of Theorem~4.2 from the work of Eisenbrand and
Shmonin~\cite{EisenbrandS08} (our interpretation is more detailed than the one
provided by Crampton et al.~\cite{CramptonGKW19}). We proceed with formally
stating our result and proving it in~Section~\ref{sec:EEFPILP}. Later, in
Section~\ref{sec:generalizing} we discuss how to extend our main result to
cover multiple further prominent fairness and efficiency concepts. In the last
section (Section~\ref{sec:conclusion}) we give conclusions. 

\section{Preliminaries}\label{sec:preliminaries}
For a positive integer~$n$, by~$[n]$ we denote the set $\{1,2, \ldots, n\}$.
We use boldface letters, like $\vex,\vey$, to represent vectors. A vector $\vex$
consisting of $n$~coordinates is said to be in~$n$~dimensions
or~$n$-dimensional and we denote its $i$-th coordinate, $i \in [n]$, by~$x_i$.
For two vectors~$\vex$ and~$\vey$ in dimensions $n_\vex$ and~$n_\vey$
respectively, vector~$(\vex, \vey)$ is a~$(n_\vex + n_\vey)$-dimensional
vector~$(x_1, \ldots, x_{n_\vex}, y_{1}, \ldots, y_{n_\vey})$. We symbolically
denote some real matrix~$A$ with $n$~rows and $m$~columns by~$A \in
\mathbb{R}^{n \times m}$. We treat $n$-dimensional vectors as matrices with
$n$~rows and $1$~column.
A \emph{polyhedron} is an intersection of half-spaces, that is, for some
dimensions~$m$ and~$n$, a polyhedron is a set~$\{ \vex \in \mathbb{R}^n : A\vex \le \veb \}$ of vectors,
for some $A \in \mathbb{R}^{m \times n}$ and $\veb \in \mathbb{R}^m$. Similarly,
assuming the same notation and defining~$\bar{A}$ and~$\bar{\veb}$ analogously, a
\emph{partially open polyhedron} is an intersection of half-spaces and open
half-spaces, that is, a set $\{ \vex \in \mathbb{R}^n : A\vex \le \veb, \,
\bar{A}\vex < \bar{\veb} \}$ of vectors.

\subsection{Allocations, Envy-Freeness, and Pareto-Efficiency}
\newcommand{\agents}{\ensuremath{\mathcal{A}}}
\newcommand{\items}{\ensuremath{\mathcal{I}}}
\newcommand{\agent}{\ensuremath{a}}
Consider a set~$\agents=\{a_1, a_2, \ldots, a_n\}$ of~$n$ agents, a
set~$\items = \{1, 2, \ldots, m\}$ of~$m$ item types with
multiplicities~$m_i$~for each item~$i \in \items$. An allocation~$\ve\pi$ is an
integral $(n \cdot m)$-dimensional vector~$\ve\pi = \left( \pi^1_{a_1}, \ldots,
\pi^1_{a_n}, \pi^2_{a_1}, \ldots, \pi^m_{a_n} \right)$, whose entries describe
for each agent how many items of each item type are allocated to the agent. For
each agent~$\agent \in \agents$,
let~$u_\agent \colon \items \rightarrow{} \mathbb{Z}$ be the agent's utility
function (in fact, utility values may be rational numbers, in which case an
equivalent problem instance with integral values can be obtained without loss
of generality by multiplying each values by the least common multiplier of the
denominators).
We assume the preferences of the agents to be additive, which means that
the utility value for a set of items is the sum of the items utility values.
Thus, we define the satisfaction of agent~$\agent \in \agents $ from
allocation~$\ve\pi$ as~$\sum_{i \in \items} u_a(i) \cdot \pi^i_{a}$; for brevity, we
slightly abuse the notation and denote it by~$u_a(\ve\pi)$.

Before we proceed, let us fix a set~$\agents$ of~$n$ agents and a set~$\items$
of~$m$~item types with multiplicities~$m_i$ for each item type~$i \in \items$.
Let~$\ve\pi$ be an allocation of the items to the agents in~$\agents$. In the
following two definitions we provide formal phrasings of envy-freeness and
Pareto-efficiency, which play a central role in our study.

\begin{definition}
\label{def:properties}%
An allocation~$\ve\pi$ of the items~$\items$ with multiplicities~$m_i$, $i \in
\items$, to agents~\agents{} is \emph{envy-free} if there is no two agents~$\agent
\in \agents$ and~$\agent' \in \agents$ such that~$u_\agent(\ve\pi) < \sum_{i \in
\items} u_\agent(i) \cdot \pi^i_{\agent'}$.
\end{definition}

\begin{definition}
An allocation~$\ve\pi$ of the items~$\items$ with multiplicities~$m_i$, $i \in
\items$, to agents~\agents{} is \emph{Pareto-dominated} if there exists another
allocation~$\ve\pi'$ (over the same sets of agents and items together with
their multiplicities) such that for every agent~$\agent \in \agents$ it holds
that~$u_\agent(\ve\pi') \geq u_\agent(\ve\pi)$ and for at least one agent the
inequality is strict. An allocation is \emph{Pareto-efficient} if it is not
Pareto-dominated.%
\end{definition}

In our work, we focus on a decision problem in which we ask whether for given
sets of agents and resources, an allocation that is simultaneously envy-free
and Pareto-efficient exists.

\defProblemQuestion{\EEFAllocationProblem}
{A set~$\agents$ of~$n$~agents, a set~$\items$ of~$m$ item types, agent utilities
$u_a\colon \items \to \mathbb{Z}$ for every $a \in \agents$, and item multiplicities~$m_i
\in \mathbb{N}$ for each $i \in \items$.}
{Is there an envy-free Pareto-efficient allocation?}

The name of the problem, standing for ``efficient envy-free'' allocation might
be misleading in the light of the fact that in the literature ``efficiency''
has multiple embodiments (besides Pareto-efficiency, perhaps the most frequent
ones are completeness or social welfare maximization). However, for clarity, we
decided to keep the name as defined by Bouveret and Lang~\cite{BL08} and then
consequently used by the follow-up works~\citep{BBN16,BredereckKKN19}.

\subsection{Parameterized Complexity}
A parameterized (decision) problem's input consists of a decision problem instance~\instance{}
and a parameter value~$k$; the task is then to decide whether $(\instance{}, k)$
is a ``yes''-instance. We say that a parameterized problem is \emph{fixed-parameter
tractable with respect to~$k$} (belongs to the class \textsf{FPT} with respect
to~$k$) if there is an algorithm deciding~$(\instance,k)$ in~$f(k) \cdot
\poly(|\instance|)$ time, where~$|\instance|$ is the size of the input and
$f(k)$ is an arbitrary computable function of parameter~$k$. Intuitively, the
exponential blow-up is then related only to the value of parameter~$k$, which
allows for efficient computation of the problem if~$k$ is small. 
The following proposition describing a relation between various functions
values will come handy later.
\begin{proposition}[{\citep[Lemma~3.10]{JonesLRSS17}}]\label{prop:Suchy}
  For every two computable functions $g\colon \mathbb{N} \to \mathbb{N}$ and $h\colon \mathbb{N} \to \mathbb{N}$ with $g(n) = o(\log (n))$, there exists a computable function $f\colon \mathbb{N} \to \mathbb{N}$ such that for every $k$ and $n$ we have
  \( 2^{g(n)h(k)} \le f(k) \cdot n \).
\end{proposition}

\subsection{Parametric Integer Programming}\label{sec:parametricILPinFixedDimension}
For a rational polyhedron $Q \subseteq \mathbb{Q}^{m+p}$, the \emph{integer
projection} of~$Q$, denoted by~$Q/\mathbb{Z}^p$, is a collection of all vectors
$\veb \in \mathbb{R}^m$ for which there exists an integral vector $\vez \in
\mathbb{Z}^p$ such that $(\veb, \vez) \in Q$. Thus, formally
\[
  Q/\mathbb{Z}^p
  :=
  \{\veb \in \mathbb{Q}^m\colon (\veb, \vez) \in Q \text{ for some } \vez \in \mathbb{Z}^p\} \,.
\]

Parametric Integer Programming (\PILP{}) is the following problem. Given a matrix
$A \in \mathbb{Q}^{m \times n}$ and a rational polyhedron $Q \subseteq
\mathbb{Q}^{m+p}$, decide if for all vectors $\veb \in \mathbb{Q}^m$ in the
integer projection of~$Q$, the system of inequalities $A \vex \le \veb$ has an
integral solution. In other words, one has to decide the validity of the
sentence
\begin{equation}\label{eq:PILP}\tag{\PILP}
 \forall \veb \in Q/\mathbb{Z}^p \ \exists \vex \in \mathbb{Z}^n \colon\quad A \vex \le \veb \,.
\end{equation}
Intuitively, \PILP{} consists of a collection of integer linear programs
defined by~$A$ and right-hand side vectors~$\veb$, where the latter ones come
from the integer projection~$Q/\mathbb{Z}^p$. The question then is whether each
of these integer linear programs has some feasible solution. The \PILP{}
problem is complete for the class $\Pi_2^p$~\citep{Stockmeyer76,Wrathall76}.

\section{Preparation for Main Result}
\label{sec:preparation}
We devote this section to describe important consequences resulting from the
work of~Eisenbrand and Shmonin~\cite[Theorem~4.1 and
Theorem~4.2]{EisenbrandS08}. Most importantly, their results
allow for efficiently solving \PILP{} subject to additional constraints. As it
will turn out, we are able to formulate \EEFAllocationProblem{} in a way
that respects these constraints. Yet, before we show the formulation
in~Section~\ref{sec:EEFPILP}, we discuss the aforementioned consequences in
detail and present them formally in Proposition~\ref{prop:solvingPILP}.  

Despite the $\Pi_2^p$-completeness of the \PILP{} problem, Eisenbrand and
Shmonin~\cite[Theorem~4.1 and Theorem~4.2]{EisenbrandS08} gave a
polynomial-time algorithm for \PILP{} for the fixed number of variables and
dimension~$n$ (their work extended the pioneering---to the best of our
knowledge---works of Kannan~\cite{Kannan90,Kannan92} on efficient algorithms
for \PILP{}).
An analysis of their algorithm leads to the following
Proposition~\ref{prop:solvingPILP}; we discuss its details afterwards.
\begin{proposition}\label{prop:solvingPILP}
  There is an algorithm deciding the sentence~\eqref{eq:PILP} in \[ f(m,n,p) \cdot \phi^{h(n)} \cdot \poly(L) \] time, where $\phi$ is the size (encoding length) of any column in~$A$, $L$ is the encoding length of the sentence and (the description of) the polyhedron~$Q$, and $f$ and $h$ are computable functions.
  Moreover, if the sentence~\eqref{eq:PILP} is not valid, then a certificate~$\veb \in Q$ is provided (i.e., $A \vex \le \veb$ has no integral solution with such a~$\veb$).
\end{proposition}

Proposition~\ref{prop:solvingPILP} essentially follows from an in-depth
analysis of a known result~\citep[Theorem~4.2]{EisenbrandS08}. A similar
investigation has also been provided by Crampton et al.~\cite{CramptonGKW19}.
However, we decided to slightly adjust it to our needs and hence we present it
in more detail. Since Proposition~\ref{prop:solvingPILP} plays an important
role in our result, we believe that discussing its argument explicitly is
important for the completeness of our paper.

In the algorithm backing~Proposition~\ref{prop:solvingPILP}, we first utilize
the Fourier--Motzkin elimination procedure to make sure that for all $\veb \in
Q$ the system $A \vex \le \veb$ has a fractional solution. If this is not the
case, then a corresponding vector~$\veb$ is reported which certifies the right-hand
side vector for which the \PILP{} sentence has no solution. Running this
procedure for all $\veb \in Q$ yielding the corresponding integer linear
programs $A \vex \le \veb$, requires solving \( f'(m,n) \) many mixed integer
linear programs in dimension~$p$. This can be done in $p^{O(p)}\poly(L)$ time
using Lenstra's celebrated result~\citep{Lenstra83} about solving integer linear
programs in bounded dimensions.

Second, we partition the polyhedron~$Q$ into~$t$ partially open
polyhedrons~$S_i$, $i \in [t]$. Due to a result by Eisenbrand and
Shmonin~\cite[Theorem~4.1]{EisenbrandS08}, the number~$t$ of partially open
polyhedra~$S_i$, $i \in [t]$, is expressed (using helper
constants~$\bar{\omega}(n)$ and~$h(n)$, which we describe below) as
\[
  t = O\left( \left( m^{2n} \phi^{n-1} \right)^{n\bar{\omega}(n)} \right) =
  f''(m,n) \cdot \phi^{h(n)} \,.
\]
Here, ${\bar{\omega}(n) = \Pi_{i = 1}^n \omega(n)}$, where $\omega(n)$ is the
constant from the flatness theorem (the current best value is~${\omega(n) =
O(n^{3/2})}$~\citep{BanaszczykLPS99}), and~$h(n) = n(n-1) \cdot
\bar{\omega}(n)$. Importantly, Eisenbrand and
Shmonin~\cite[Theorem~4.1]{EisenbrandS08} show that each~$S_i$, $i \in [t]$, is
an integer projection of some partially open polyhedron~$S_i'$, that is $S_i =
S_i' / \mathbb{Z}^{\ell_i}$; additionally they show that ${\ell_i =
O(\bar{\omega}(n))}$, $i \in [t]$.
Lastly, the result of Eisenbrand and Shmonin~\cite[Theorem~4.1]{EisenbrandS08}
gives, for each $i \in [t]$, a collection of~$k_i = f'''(n)$ specific
transformations~$T_{ij}$, for~$j \in [k_i]$. The transformations are very
specific in the sense that for each $\veb \in S_i$ there is an integral point in
the polyhedron~${P_{\veb} := \left\{ \vex : A \vex \le \veb \right\}}$ if and
only if $T_{ij}(\veb) \in P_\veb$ for some $j \in [k_i]$.
The negation of this condition can be verified using a mixed integer linear program for each~$i
\in [t]$; such an ILP has~$(k_i+1)n+\ell_i+p$ integral variables.
It holds that if the input sentence~\eqref{eq:PILP} is not valid, then one of the
above mixed ILPs is feasible; thus, again, providing the claimed certificate~$\veb$.
Carefully inspecting the two parts of the above-sketched algorithm reveals
that it runs in the requested \( f(m,n,p) \cdot \phi^{h(n)} \cdot
\poly(L)\)~time.

\section{Finding EEF--Allocations via PILP}
\label{sec:EEFPILP}
The interpretation of Theorem~4.2 of Eisenbrand and
Shmonin~\cite{EisenbrandS08} presented in Section~\ref{sec:preliminaries}
contains an important bit. Specifically, we observed that it is possible
to derive a certificate of infeasibility of a given \PILP{} sentence.
This inspired us to consider the following reasoning, which we employ to derive
our result about finding envy-free and Pareto-efficient allocations. Instead of
focusing directly on~\EEFAllocationProblem{}, we decided to work with
the~complementary problem. This way, by obtaining the certificate
of infeasibility for the complementary problem, we in fact get a (membership)
certificate for the original problem. In more details, we think of a problem of
deciding whether ``every envy-free allocation is Pareto-dominated.'' If such a
sentence is invalid, then a certificate proving it is  an envy-free allocation
that cannot be Pareto-dominated. It is worth pointing out that due to the
certificate, we do not only answer the question posed
by~\EEFAllocationProblem{} but we also find an envy-free and Pareto-efficient
allocation, which makes our approach constructive.

The method described above leads us to the main contribution of our work, which
strengthens Corollary~5 of Bredereck et al.~\cite{BredereckKKN19} about
fixed-parameter tractability of~\EEFAllocationProblem{} with respect to the
combined parameter ``number of agents plus number of items.'' Therein, the
authors devise the negation of~\EEFAllocationProblem{} in a similar spirit to
ours (however, their approach is fundamentally different as it is based on
analyzing a collection of improving steps among which none can be added to
improve a given allocation) employing the big-M method to do so. 
We avoid this method, which (as used in the mentioned paper) forces a unary
encoding of the input item multiplicities and utility values, arriving at our
Theorem~\ref{thm:EEFAllocationProblemFPTforSmallUtilities}, which offers the
same computational complexity guarantees but \emph{does not} require the unary
encoding of the discussed input elements.

\begin{theorem}\label{thm:EEFAllocationProblemFPTforSmallUtilities}
  Let~$\instance$ be an instance of the \EEFAllocationProblem problem with the
  maximum input utility value~$u_{\max} = 2^{o(\log|\instance|)}$.
  Then, there is an algorithm that decides~$\instance$ in $f(m+n) \cdot
  \poly(|\instance|)$ time, for some computable function~$f \colon \mathbb{N}
  \to \mathbb{N}$ and~$|\instance|$ being the size of~$\instance$.
\end{theorem}

Before we proceed with
proving~Theorem~\ref{thm:EEFAllocationProblemFPTforSmallUtilities} in the
following Section~\ref{sec:proving}, we remark that our technique also applies
to other variants of \EEFAllocationProblem where we replace envy-freeness or
Pareto-efficiency with related concepts. We devote a separate section
(Section~\ref{sec:generalizing}) to a detailed discussion about these
additional applications.

\subsection{Proving the result}\label{sec:proving}

Employing Proposition~\ref{prop:solvingPILP}, we now show how to efficiently
solve the \EEFAllocationProblem problem for the (combined) parameter ``number
of agents plus number item types,'' obtaining a proof
of~Theorem~\ref{thm:EEFAllocationProblemFPTforSmallUtilities}. From now on, we
fix a set~$\agents$ of~$n$ agents and a set~$\items = \{1, 2, \ldots, m\}$
of~$m$~item types with multiplicities~$m_i$, $i \in \items$.

As already discussed, we show the \textsf{FPT}-membership
of~\EEFAllocationProblem for the parameter $n+m$ by constructing
a~\PILP{} sentence deciding whether every envy-free allocation of a
given collection of items is dominated by some other allocation. The high-level
idea is as follows. We first construct the~\PILP{} sentence (which essentially
corresponds to the matrix~$A$ in~Formula~\eqref{eq:PILP}) assuming that we
have a polyhedron~$Q$ that describes all envy-free allocations. Then we show
how to construct the polyhedron~$Q$ such that it meets our assumptions. (In fact, the
polyhedron also contains additional technical parts needed to represent that
there is an allocation that dominates some allocation from the polyhedron.) Eventually,
we use the results from~Proposition~\ref{prop:solvingPILP}.
Starting our proof with assuming that we have polyhedron~$Q$ and showing its
construction later is due to the fact that the former will develop our
intuition how the polyhedron~$Q$ should look like. Before we go ahead with the
proof, we recall that an allocation~$\vex$ consists of entries~$x_i^a$, for
each agent~$a \in \agents$ and item type~$i \in \items$, with the meaning ``we
give $x^i_a$~items of item type~$i$ to agent~$a$.''

\paragraph{Describing Domination of Allocation with Matrix~$A$.}
Let us assume such a polyhedron~$Q$ that we have~$(\veb,\vez) \in Q$,
where~$\vez$ is an allocation (we do not discuss~$\veb$ as it is still to be
defined in the next point of the proof where we construct a proper~$Q$).
Our aim is to design a matrix~$A$ such that $A \vex \le \veb$ if and only
if~$\vex$ is an allocation that dominates~$\vez$. We first focus on constraints
enforcing that~$\vex$ is a proper allocation (not necessarily allocating all
items to the agents; this will be guaranteed later due to
the requirement of Pareto-efficiency).
\begin{align}
  \sum_{a \in \agents} x^a_i &\le m_i     & \forall i \in \items                             \label{eq:xIsAllocation:1} \\
  x^a_i &\ge 0                      & \forall a \in \agents, \forall i \in \items            \label{eq:xIsAllocation:2}
\end{align}
Condition~\eqref{eq:xIsAllocation:1} ensures that $\vex$ does not allocate
``more items than available,'' while Condition~\eqref{eq:xIsAllocation:2}
guarantees that each agent~$a \in \agents$ is allocated a non-negative number of
items by~$\vex$. It is now not hard to see that~$\vex$ satisfies
Conditions~\eqref{eq:xIsAllocation:1} and~\eqref{eq:xIsAllocation:2} if and
only if~$\vex$ is a valid allocation.

Thus, it remains to model that~$\vex$ Pareto-dominates~$\vez$.
One can do so with the following system of inequalities. Note that on the
right-hand side we use the (entries of the) vector~$\vez$; we do so for brevity
of our proof. In the final~\PILP{} sentence the right-hand side must be defined
by~$\veb$ and we will indeed use the insights from the following inequalities
to define~$\veb$ (as a part of defining~$Q$) in the next step of our proof.
\begin{align}
  \sum_{i \in \items} u_a(i) \cdot x^a_i &\ge \sum_{i \in \items} u_a(i) \cdot z^a_i                                    & \forall a \in \agents              \label{eq:xDominatesY:1} \\
  \sum_{a \in \agents} \sum_{i \in \items} u_a(i) \cdot x^a_i &\ge 1 + \sum_{a \in \agents} \sum_{i \in \items} u_a(i) \cdot z^a_i                                          \label{eq:xDominatesY:2}
\end{align}
The system of inequalities above guarantees that~$\vex$ dominates~$\vez$ if and
only if it satisfies Conditions~\eqref{eq:xDominatesY:1} and~\eqref{eq:xDominatesY:2}.
Note that Condition~\eqref{eq:xDominatesY:1} ensures that the total utility
of each agent~$a \in \agents$ in allocation~$\vex$ is at least as good as
that of agent~$a$ in allocation~$\vez$.
Furthermore, given the above, the condition described by
Inequality~\eqref{eq:xDominatesY:2} ensures that there is at least one
agent~$a \in \agents$ for whom it holds that \(\sum_{i \in \items} u_a(i) \cdot x^a_i > \sum_{i
\in \items} u_a(i) \cdot z^a_i\), that is, whose utility is greater in
allocation~$\vex$ than that in~$\vez$.

\paragraph{The Polyhedron~$Q$.}
We now aim at designing an appropriate polyhedron~$Q$, existence of which we
(only) assumed in the first step. Given the above discussion and
Conditions~\eqref{eq:xIsAllocation:1}--\eqref{eq:xDominatesY:2}, we have that
the claimed $\veb$ is in dimension~$m + mn + n + 1$, that is $\veb \in
\mathbb{Q}^{m + mn + n + 1}$. Indeed, the summands in $\veb$'s dimension
expression come directly from the numbers of inequalities in, respectively,
Conditions~\eqref{eq:xIsAllocation:1}--\eqref{eq:xDominatesY:2}. Since we
assumed that~$\vez$ is an allocation, we have~$\vez \in \mathbb{Z}^{mn}$ by
definition. Overall, it must hold that $Q \subseteq \mathbb{Q}^{m + 2mn + n +
1}$.

Let us now split the vector~$\veb = (\veb_1, \veb_2, \veb_3, b_4)$ according to
Conditions~\eqref{eq:xIsAllocation:1}--\eqref{eq:xDominatesY:2} above---that
is, $\veb_1$ is the vector of right-hand sides coming from
Condition~\eqref{eq:xIsAllocation:1} and so forth. Based on the first two
subject conditions, we thus have
\begin{align}
  \veb_1 = \vem \,, \label{eq:conditionForB:1a} \\
  \veb_2 = \vezero \,, \label{eq:conditionForB:1b}
\end{align}
where $\vem$ is the vector of item multiplicities. Clearly, if we now use the
above-defined~$\veb_1$ and~$\veb_2$ substituting the right-hand sides of,
respectively, Conditions~\eqref{eq:xIsAllocation:1}
and~\eqref{eq:xIsAllocation:2}, the meaning of
Conditions~\eqref{eq:xIsAllocation:1} and~\eqref{eq:xIsAllocation:2} stays
intact. More precisely, both conditions still encode the fact that~$\vex$ is an
allocation.

We proceed with constructing vector~$\veb_3$ and the value of~$b_4$. To
achieve this, we first ensure that~$\vez$ is an envy-free allocation and then
derive~$\veb_3$ and~$b_4$ from this analysis.
The following conditions ensure that~$\vez$ is an envy-free allocation.
\begin{align}
  \sum_{a \in \agents} z^a_i &\le m_i                                                 & \forall i \in \items                             \label{eq:zIsAllocation:1} \\
  z^a_i &\ge 0                                                                  & \forall a \in \agents, \forall i \in \items            \label{eq:zIsAllocation:2} \\
  \sum_{i \in \items} u_a(i) \cdot z^a_i &\ge \sum_{i \in \items} u_a(i) \cdot z^{a'}_i
				    & \forall a,a' \in \agents
				    \label{eq:zIsEnvyFree}
\end{align}
Conditions~\eqref{eq:zIsAllocation:1} and~\eqref{eq:zIsAllocation:2} ensure
that~$\vez$ encodes an allocation. These expressions and hence the argument are
analogous to those of Conditions~\eqref{eq:xIsAllocation:1}
and~\eqref{eq:xIsAllocation:2} for~$\vex$.
Further, Condition~\eqref{eq:zIsEnvyFree} ensures that~$\vez$ is envy-free,
since the left-hand side is the total satisfaction of agent~$a$ (under
allocation~$\vez$) and the right-hand side is the total value of the bundle
of~$a'$ viewed via the utility function of agent~$a$ (that is, the satisfaction
of~$a$ if she got the bundle that~$a'$ gets under allocation~$\vez$).
At the moment, intuitively,
Conditions~\eqref{eq:conditionForB:1a}--\eqref{eq:zIsEnvyFree} describe the ``part''
of polyhedron~$Q$ that defines~$\veb_1$, $\veb_2$, and~$\vez$. What remains, is
to define the remaining~$\veb_3$ and~$v_4$ in a way that we can use them as the
right-hand sides of~Conditions~\eqref{eq:xDominatesY:2}
and~\eqref{eq:xDominatesY:1}, respectively. We can do so by binding~$\vez$
to~$(\veb_3,b_4)$ as follows, thus obtaining the final two expressions describing
polyhedron~$Q$.
\begin{align}
  \sum_{i \in \items} u_a(i) \cdot z^a_i &= b_3^a                &\forall a \in \agents \label{eq:conditionForB:2} \\
  \sum_{a \in \agents} \sum_{i \in \items} u_a(i) \cdot z^a_i &= b_4                    \label{eq:conditionForB:3}
\end{align}
Observe that the left-hand side of Condition~\eqref{eq:conditionForB:2} is exactly the
right-hand side of~\eqref{eq:xDominatesY:1}. Similarly, the right-hand side
of~\eqref{eq:conditionForB:3} contains exactly (up to the constant $1$) the
right-hand side of Condition~\eqref{eq:xDominatesY:2}.
Consequently, we can replace the right-hand sides of Conditions~\eqref{eq:xDominatesY:1}
and~\eqref{eq:xDominatesY:2} with the right-hand sides
of Conditions~\eqref{eq:conditionForB:2} and~\eqref{eq:conditionForB:3} while
keeping the meaning of the latter unchanged. Observing that in this last step
we defined the whole~$\veb$ in a way that allows us using~$\veb$ in the
right-hand sides of
Conditions~\eqref{eq:xIsAllocation:1}--\eqref{eq:xDominatesY:2}, we arrive at
the next lemma, which summarizes (and follows) from the above discussion.

\begin{lemma}\label{lem:polyhedron}
  Let~$Q \subseteq \mathbb{Q}^{m + 2mn + n + 1}$ be a polyhedron defined by the conditions~\eqref{eq:conditionForB:1a}--\eqref{eq:conditionForB:3}.
  Then, $(\veb, \vez) \in Q$ if and only if
  \begin{itemize}[leftmargin=5ex]
    \item $\vez$ is an envy-free allocation of the items described by~$\vem$,
    \item $\veb$ is the vector of right-hand sides of
	    Conditions~\eqref{eq:xIsAllocation:1}--\eqref{eq:xDominatesY:2}.
  \end{itemize}
\end{lemma}

We remark that the fact that Conditions~\eqref{eq:zIsEnvyFree} and
\eqref{eq:conditionForB:3} are presented in a way that the right-hand side is
not a constant is not important in the light of the definition of~$Q$
from~Lemma~\ref{lem:polyhedron}. Clearly, to obtain a constant on the
right-hand sides it is enough to substract the right-hand side from both sides
starting from the expressions presented in Conditions~\eqref{eq:zIsEnvyFree} and
\eqref{eq:conditionForB:3}.

\paragraph{Using Proposition~\ref{prop:solvingPILP}.}
Having described how to construct the parametric ILP
representing~\EEFAllocationProblem{}, we finish the proof of
Theorem~\ref{thm:EEFAllocationProblemFPTforSmallUtilities} by applying
Proposition~\ref{prop:solvingPILP}. More specifically, for a given
instance~\instance{} of the \EEFAllocationProblem problem, we construct
matrix~$A$ and polyhedron~$Q$ as described earlier and directly build a
parametric PILP instance~$\instance'$ out of them.  Then we run the algorithm
from Proposition~\ref{prop:solvingPILP} on instance~$\instance'$.  If the
algorithm returns ``yes,'' then for every envy-free allocation there exists one
that dominates it, so the answer to the original instance~$\instance$ is
``no.'' In the opposite case, we know that~$\instance$ admits some Pareto-efficient
envy-free allocation~$\vex$, so we output ``yes'' as an answer to~$\instance$.
Moreover, due to the fact that Proposition~\ref{prop:solvingPILP} guarantees
returning a certificate, the ``no''-certificate computed by the algorithm is in
fact the envy-free Pareto-efficient allocation~$\vex$.

It remains to analyze the running time of the invocation of the algorithm
from~Proposition~\ref{prop:solvingPILP} on the constructed
instance~$\instance'$.
In the presented model, described
by~\eqref{eq:xIsAllocation:1}--\eqref{eq:conditionForB:3}, forming
instance~$\instance'$, the dimension of~$\vex$ is $m \cdot n$, where~$n$ is the
number of agents in~$\instance$ and~$m$ is the number of item types. Hence, the
value of parameter~$p$ from Proposition~\ref{prop:solvingPILP} is~$p = m \cdot
n$. It remains to estimate the parameter~$\phi$ thereof. Recall that~$\phi$ is
the maximum encoding length of a column in~$A$, which is, in our case, the
matrix of left-hand sides
in Conditions~\eqref{eq:xIsAllocation:1}--\eqref{eq:xDominatesY:2}.
The columns of the matrix~$A$ are vectors of length $mn + 2m + 1$---this length
is equal to the number of constraints (inequalities) required to implement
these conditions. Hence, there are $m(n+2)$ many delimiter symbols in the
encoding of a single column.
Recall that each such column corresponds to a pair, a single agent~$a \in
\agents$ and a single item~$i \in \items$, and let us fix some pair~$(a, i)$.
So, in the column of~$(a, i)$, there are~$2$ ones, one coming from
Condition~\eqref{eq:xIsAllocation:1} and one from
Condition~\eqref{eq:xIsAllocation:2}. In addition to this, there are~$2$
numbers, both equal to~$u_a(i)$. Since we assumed a binary encoding, that is $u_a(i) =
2^{o(\log|\mathcal{I}|)}$, we overall obtain the encoding length
$2^{o(\log|\instance|) + 1} + m(n+2)$ of a single column, which, after dropping
the asymptotically irrelevant terms, gives $\phi = 2^{o(\log|\instance|)}$.
Due to Proposition~\ref{prop:Suchy}, we thus get that there is a
function~$\hat{f}(n)$ such that $\phi^{h(n)} \le \hat{f}(n) \cdot |I|$.
Applying this value for~$\phi^(h(n))$, together with the one for $p$ shown
earlier, proves that the algorithm from Proposition~\ref{prop:solvingPILP} runs
in the running time required to show fixed-parameter tractability of
\EEFAllocationProblem{} with respect to the parameter~$m+n$.

\section{Generalizing Our Approach}\label{sec:generalizing}
Envy-freeness is an appealing yet demanding concept. Consider a very simple
example of two agents desiring a single item. Already in this situation an
allocation that allocates the item cannot be envy-free. Hence, there is no
nontrivial envy-free allocation of items (recall that an empty allocation is
always envy-free).

The experimental results of~Bredereck et~al.~\cite{BFKKR21} give empirical evidence that
non-existence of envy-free and Pareto-efficient allocations poses a real threat
to applicability of these concepts in real-world instances. The authors show
that there were no envy-free and Pareto-efficient allocations for 63\% of the
instances in their dataset from spliddit.org. The observed phenomenon clearly
motivates the need for general approaches. In practice, in the case of a
scenario with no envy-free and Pareto-efficient allocation, a reasonable
algorithm should not only report the non-existence but also offer a
possibly-best alternative allocation, which yields weaker desiderata. The
current state of the art in the form of both, an extensive literature on
envy-freeness relaxations (see our Related Work section for the references) and
general frameworks presented by~Bredereck et~al.~\cite{BFKKR21, BredereckKKN19}
strongly suggest that providing generalizable results is of high value. 

Our method meets this criterion and can be used with numerous other problem
variants that aim at finding efficient fair allocations. Indeed, it turns out
that our technique can be applied to the \EFProblem
problem~\citep{BredereckKKN19}, which is a more general variant of the
\EEFAllocationProblem where Pareto-efficiency is replaced by some efficiency
notion~\ensuremath{\mathcal{E}} and envy-freeness is replaced by some fairness
notion~\ensuremath{\mathcal{F}}. Formally, the problem, as defined by~Bredereck
et~al.~\cite{BredereckKKN19}, is as follows.

\defProblemQuestion{\EFProblem}
 {A set of agents $A$, a set of item types $I$, agent utilities $u_a\colon I \to \mathbb{Z}$ for every $a \in A$, and item multiplicities $m_i \in \mathbb{N}$ for $i \in I$.}
 {Is there an $\mathcal{F}$-free allocation which is $\mathcal{E}$-efficient.}

In fact, our approach can be used to show fixed-parameter tractability 
of the above problem with respect to the parameterization by the number~$n$ of agents
plus the number~$m$ of item types for various efficiency and fairness notions.
Besides relaxed notions of Pareto-efficiency (e.g.,\ where one only cares about
being dominated by allocations to some extent similar to the to-be-dominated
one) or relaxed envy-freeness such as EF1~\citep{BKV18,CKMPSW16,LMMS04} or
EFX~\citep{CKMPSW16,PM18}, our approach can also deal with generalizations of
the concepts of Pareto-optimality such as such as group
Pareto-efficiency~\citep{AW18} or generalizations of envy-freeness such as graph
envy-freeness~\citep{BKN22}. Additionally, our method is adaptable to further
somewhat related fairness concepts such as MaxiMinShare~\citep{Bud11,PW14} or a
basic efficiency concept completeness, which only requires that all resources
are allocated.

Summarizing, with our technique we can show that \EFProblem is fixed-parameter
tractable for parameter $n+m$ even if item multiplicities and utilities are
binary encoded when
\begin{itemize}[leftmargin=5ex]
  \item $\mathcal{E}$ is a combination of (graph/group) Pareto-efficiency or completeness, and
  \item $\mathcal{F}$ is a combination of (graph/group) EF, (graph) EF1,
        (graph) EFX, Maxi\-Min, or MaxiMinShare.
\end{itemize}
To avoid repetitiveness, we refer to the work of
Bredereck~et~al.~\cite{BredereckKKN19} on how to model these notions within the
ILP framework.

\section{Conclusion}
\label{sec:conclusion}

We described a somewhat new usage of Parametric ILPs in fixed dimension in the
design of parameterized algorithms, enabling to improve a
previous fixed-parameter tractability result.
To the best of our knowledge, we are the first to model (and solve) the
negation of a given instance to obtain a solution to the original in the
context of parameterized complexity. Thus, we believe to have contributed to
the, recently gaining increased attention (see, for example, a survey by
Gaven\v{c}iak et al.~\cite{GKK22}), understanding of how the theory of integer
(linear) programming impacts the theory of parameterized complexity. We hope
our approach leads to further new results in parameterized algorithms, including
applications beyond social choice. 

Our work also brings up new challenges and highlights the importance of some
yet unexplored research directions, mostly in the area of empirical study of
efficient and fair allocations of indivisible items.

First of all, given a practically applicable implementation~\citep{BFKKR21} of
the approach of~Bredereck et al.~\cite{BredereckKKN19}, it appears valuable to
pursue an empirical study of our approach as well. It is not uncommon that
algorithms with appealing (worst-case) computational complexity guarantees do
not perform that well when applied to real-life instances. Hence, designing an
implementation of our method and comparing it against the existing methods of
computing efficient and fair allocations is a necessary step in judging the
usability of our study in practice.

Performing computational experiments is a natural next step to gain additional
insights into the problem nature (like, a sharp phase transition in the
existence of efficient envy-free allocations reported by~Dickerson et
al.~\cite{DGKPS14}). Offering a next tool in the algorithmic toolbox for
seeking fair allocations, we also highlight the need for further efforts
towards obtaining realistic data or, at least, designing diversified synthetic
models of generating allocation instances. By now, to the best of our
knowledge, except for the relatively small dataset of real-world data from the
website spliddit.org~\citep{GP14} and two very simple synthetic models
by~Dickerson et al.~\cite{DGKPS14}, such data is lacking. Our method might not
only turn out to be useful in spotting new phenomena of fair allocation
instances, but might also well complement other existing methods to form a
robust framework for finding fair and efficient allocations.
 
\section*{Acknowledgments}
We are thankful to the ECAI '23 reviewers for their helpful comments.
This work was started when all authors were with TU~Berlin.
Du\v{s}an Knop acknowledges the support of the Czech Science Foundation Grant
No. 22-19557S.
Andrzej Kaczmarczyk and Robert Bredereck were supported by the DFG, project
AFFA (BR 5207/1 and NI 369/15).
This project has received funding from the European Research Council
(ERC) under the European Union's Horizon 2020 research and innovation
programme (grant agreement No 101002854).

\begin{center}
  \includegraphics[width=3cm]{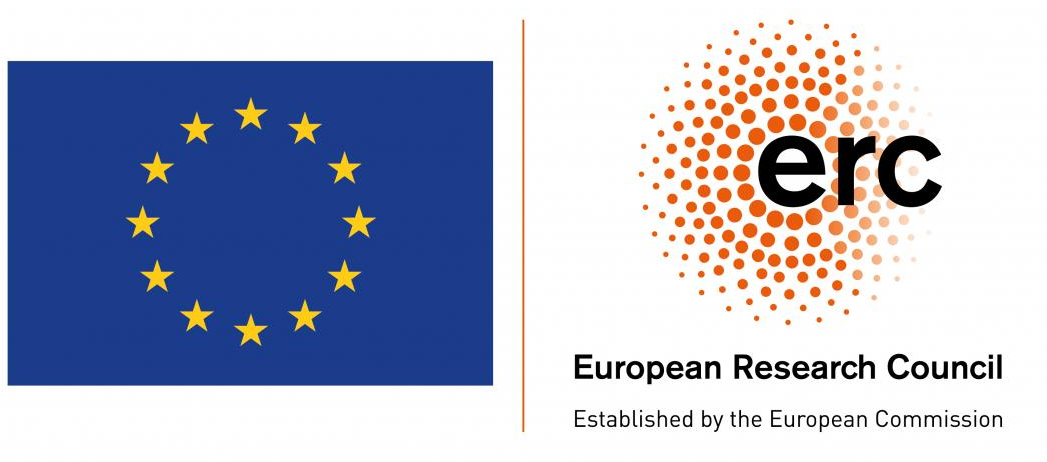}
\end{center}

\bibliographystyle{plainnat}
\bibliography{envyfree}

\end{document}